\documentclass[preprint,12pt]{elsarticle}

\usepackage{epsfig}

\usepackage{amssymb}

\usepackage{lineno}

\journal{Nuclear Instruments \& Methods in Physics Research}
\begin{document}
\def\met {\kern -0.1em{m}}
\newcommand \spec {spectrometer}
\newcommand \pcal {photon calorimeter}
\def\mm {\kern -0.1em{mm}}
\newcommand \up {up}
\newcommand \dn {down}
\newcommand \z {z} 
\newcommand \x {x}
\newcommand \y {y}
\def\kHz {\kern -0.1em{kHz}}
\def\Hz {\kern -0.1em{Hz}}
\newcommand \record {environmental record}
\newcommand \records {environmental records}
\def\gev {\kern -0.1em {Ge\kern -0.1em{V}}}
\def\cm {\kern -0.1em{cm}}
\def\mev {\kern -0.1em{MeV}}
\newcommand \Fig {Figure}
\newcommand \rmscut {rms-cut}
\newcommand \Eq {Eqn.}
\def\nbi {\kern -0.1em{nb^{-1}}}
\def\pbi {\kern -0.1em{pb$^{-1}$}}
\newcommand \sect {section}
\begin{frontmatter}


\title{Measurement of the Luminosity in the ZEUS Experiment at HERA II}

\author[UST]{L. Adamczyk}
\author[INP]{J.~Andruszkow\fnref{fn1}}
\author[UST]{T.~Bold}
\author[INP]{P.~Borzemski\fnref{fn2}}
\author[MPI]{C.~Buettner} 
\author[MPI]{A.~Caldwell}
\author[INP]{J.~Chwastowski}
\author[INP]{W. Daniluk}
\author[DESY,NCPHEP]{V. Drugakov}
\author[INP]{A. Eskreys\fnref{fn2}}
\author[INP]{J. Figiel}
\author[INP]{A. Galas} 
\author[INP]{M. Gil\fnref{fn3}} 
\author[Columbia]{M. Helbich}
\author[DESY]{F. Januschek}
\author[INP]{P. Jurkiewicz}
\author[UST]{D. Kisielewska}
\author[DESY]{U.~Klein\fnref{fn4}}
\author[INP]{A. Kotarba}
\author[DESY,BTU]{W. Lohmann\corref{cor1}}
\ead{wolfgang.lohmann@desy.de}
\author[Columbia]{Y. Ning}
\author[INP]{K. Oliwa}
\author[INP]{K. Olkiewicz}
\author[Columbia]{S. Paganis\fnref{fn5}}
\author[INP]{J. Pieron}
\author[UST]{M. Przybycien}
\author[Columbia]{Z. Ren}
\author[INP]{W. Ruchlewicz\fnref{fn1}}
\author[MPI]{W. Schmidke\fnref{fn6}} 
\author[DESY]{U. Schneekloth}
\author[Columbia]{F. Sciulli}
\author[INP]{P. Stopa}
\author[UHH]{J. Sztuk-Dambietz\fnref{fn7}}
\author[UST]{L. Suszycki}
\author[MPI]{J. Sutiak}
\author[INP]{W. Wierba}
\author[INP]{L. Zawiejski}
\cortext[cor1]{Corresponding author}
\fntext[fn1]{now at Motorola software center, Krakow}
\fntext[fn2]{deceased}
\fntext[fn3]{now at Comarch, Krakow}
\fntext[fn4]{now at University of Liverpool}
\fntext[fn5]{now at The University of Sheffield}
\fntext[fn6]{now at Brookhaven National Laboratory}
\fntext[fn7]{now at European X-Ray Free-Electron Laser Facility GmbH, Hamburg}
\address[UST]     {AGH University of Science and Technology, Cracow, Poland}
\address[INP]{ The Henryk Niewodniczanski Institute of Nuclear Physics,
   Polish Academy of Sciences, Cracow , Poland}
\address[MPI]   {Max-Planck-Institut fuer Physik, M\"unchen, Germany}
\address[DESY]  {Deutsches Elektronen Synchrotron, Hamburg, Germany}
\address[NCPHEP] {NCPHEP, Minsk, Belarus}   
\address[Columbia]{Columbia University, New York, USA}
\address[BTU] {Brandenburgische Technische Universit\"at, Cottbus, Germany}
\address[UHH] {Hamburg University, II Institute of Exp. Physics, Hamburg, Germany}

\begin{abstract}
 The luminosity in the ZEUS detector 
 was  measured using photons from electron bremsstrahlung.
 In 2001 the HERA collider was upgraded for operation
 at higher luminosity.
 At the same time the luminosity-measuring system
 of the ZEUS experiment was modified to tackle the expected higher photon rate and 
 synchrotron radiation.
 The existing 
 lead-scintillator calorimeter was equipped with radiation hard scintillator tiles and
 shielded against synchrotron radiation.
 In addition, a magnetic spectrometer was installed to measure
 the
 luminosity independently using photons converted in the beam-pipe exit window.
 The redundancy provided a reliable and 
 robust luminosity determination with a systematic uncertainty of 1.7\%.
 The experimental setup, the techniques used for luminosity 
 determination and the estimate of the
 systematic uncertainty are reported.

\end{abstract}

\begin{keyword}
luminosity measurement, ZEUS experiment


\end{keyword}

\end{frontmatter}



\section{Introduction}
\label{sec-intr}
\setcounter{footnote}{0}
At the HERA storage rings electrons\footnote{In this paper, ``electron'' is meant to specify both electron or positron.} and protons have been accelerated to energies of 
27.5~\gev~and 920~\gev, respectively, and brought 
to collisions inside two experiments, ZEUS and H1.
Data taken with these experiments provided us with precise measurements of quark and 
gluon distributions in the proton and allowed precision tests of 
Quantum Chromodynamics as 
the theory of the strong force at small distances.  
The luminosity is a key parameter of the storage ring determining its physics potential in 
terms of event statistics in the processes of interest
characterised by a certain cross section. 
The precise measurement of the 
luminosity in the experiments is of crucial importance, since
the luminosity uncertainty translates directly to the uncertainty of the
cross section and hence to the measurement of quark and 
gluon distributions.

The luminosity is measured in electron-proton scattering 
using the rate of high energy photons from 
the bremsstrahlung process, $ep \to ep\gamma$. This is a pure QED process, 
which has a high rate and a precisely calculable cross section.
In the first phase of HERA a photon calorimeter~\cite{pcal} was used to measure the rate 
of bremsstrahlung photons.
This calorimeter was positioned in the HERA tunnel
about 100 \met{} downstream of the ZEUS experiment, where 
the electron and proton beams were 
separated.  Bremsstrahlung photons moved through a straight vacuum chamber 
and an aluminum
exit window before entering the \pcal.  

After the completion of an upgrade
program the HERA accelerator restarted operations in 2001.
The luminosity was steadily increased reaching a factor of 5 times the one of the first HERA phase.
The major changes were stronger beam focusing and slightly larger beam currents.
The 
physics 
program was extended to 
processes with lower cross section e.g., to explore the heavy flavour content 
of the proton or to probe weak interactions using in addition electron beam 
polarisation. 

However, the earlier beam separation and stronger focusing of 
the beam led to more synchrotron radiation
and a larger rate of bremsstrahlung events, the latter 
resulting in a pile-up of more than one photon per bunch crossing in the  \pcal.
To reduce the impact of synchrotron radiation on the  \pcal, the thickness of the carbon
absorber blocks installed in front of the \pcal~was increased from 2 to 4 radiation lengths.
A second luminometer, the \spec~\cite{spec}, was installed outside 
the synchrotron radiation cone, to measure the luminosity
using a different detector technique. 
Bremsstrahlung photons were measured after they converted to 
electron pairs in the exit window of the photon beam 
pipe. The converted electron 
pair was spatially split by the magnetic field of a dipole, 
and both particles were individually measured by two electromagnetic calorimeters. 
The calorimeters were placed vertically at sufficiently large distance
from the photon beam not to be affected 
by the direct synchrotron radiation and unconverted bremsstrahlung photons. The large flux of 
low energy electrons from synchrotron photon interactions in 
material upstream of the dipole magnet was swept away from the calorimeters.
The data rate in the~\spec~was, in comparison to~\pcal, considerably reduced due to the 
conversion probability of about 9\% in the beam-pipe
exit window and a higher threshold on the photon energy. 
The fraction of bunch crossings with more than one photon
creating depositions in the~\spec~approached only a  
few percent at the highest instantaneous 
luminosity.

Utilising two independent luminosity measurements allowed a permanent comparison, 
which was  
important for tracing back in real time detector malfunctions and to reduce
systematic uncertainties.  

In this paper, we report on the operation of the~\pcal~and~\spec~in the data taking periods
after the upgrade, 
describe the steps to obtain the luminosity from the raw 
data, and evaluate systematic uncertainties of the luminosity measurement.
Since the spectrometer was a new device, more detailed studies are reported.

\section{Bremsstrahlung and Luminosity Determination}
\label{sec-spec-lumi}
The energy spectrum of 
bremsstrahlung photons 
in the process
\begin{equation} \label{eq-brems}
ep \to e p \gamma
\end{equation}
is given  
by Bethe and Heitler~\cite{Bethe-Heitler} 
in Born approximation neglecting the finite size of the proton as:
\begin{equation} \label{eq-bh}
\frac{d\sigma}{dE_{\gamma}} = 4 \alpha r^{2}_{e} \frac{E_{e}^{\prime}}{E_{\gamma}E_{e}} \left( \frac{E_{e}}{E_{e}^{\prime}} + 
\frac{E_{e}^{\prime}}{E_{e}} - \frac{2}{3} \right) \left( \ln\frac{4E_{p}E_{e}E_{e}^{\prime}}{m_{p}m_{e}E_{\gamma}} - \frac{1}{2} \right),
\end{equation}
where $\alpha$ is the fine structure constant,
$r_{e}$ the classical electron radius,
$E_{\gamma}$ the energy of the radiated photon,
$E_{p}$ and $E_{e}$ are the proton and electron beam energies,
$E_{e}^{\prime}$ is the scattered electron energy, and
$m_{p}$ and $m_{e}$ are the proton and electron masses. For illustration the photon energy spectrum
is shown in \Fig{~\ref{fig-energy} for an electron beam energy of 27.5~\gev.
Recently one-loop QED radiative corrections have been calculated for the 
process~(\ref{eq-brems})~\cite{makarenko}. Their impact on the cross section~(\ref{eq-bh})
and photon angular distribution 
is considered in section~\ref{sec-theory}.
\begin{figure}[!htb]
\centering
\includegraphics[width=0.7\textwidth]{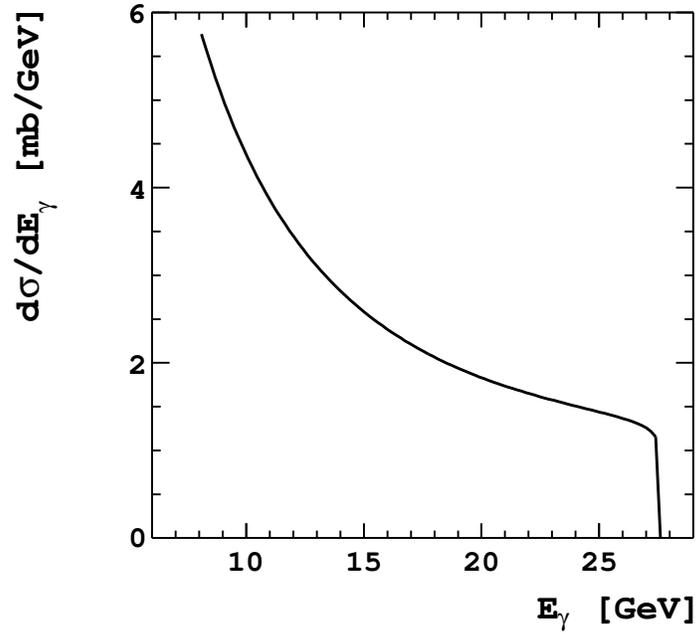}
\caption{The energy spectrum of bremsstrahlung photons obtained from \Eq(\ref{eq-bh}).
}
\label{fig-energy}
\end{figure}

The integrated luminosity is calculated for a certain data taking period as:
\begin{equation} \label{eq-lumi}
L = \frac{N_{\gamma}}{A\sigma},
\end{equation}
where $N_{\gamma}$ is the number of
bremsstrahlung photons selected in a certain energy range over the period, 
and $\sigma$ is the bremsstrahlung cross-section 
integrated over the same energy range.
The acceptance, A, is 
the probability for a bremsstrahlung photon created in 
the $ep$ interaction region to be observed in the luminometer. To estimate the acceptance
of the~\pcal~and the~\spec~detailed Monte Carlo 
simulations of both devices were done as described in section~\ref{monte-carlo}}.

\section{Photon Calorimeter and Spectrometer}
\label{sec-cal-spec}

The performances of the photon calorimeter and the spectrometer 
were studied in test-beams. The results are 
described in detail in 
Refs.~\cite{pcal,spec}.

Both devices were installed about 100~\met{} downstream from the 
nominal interaction point, IP, in the direction of the electron beam, as 
sketched in \Fig{~\ref{fig-layout}.
\begin{figure}[!htb]
\centering
\includegraphics[width=\textwidth]{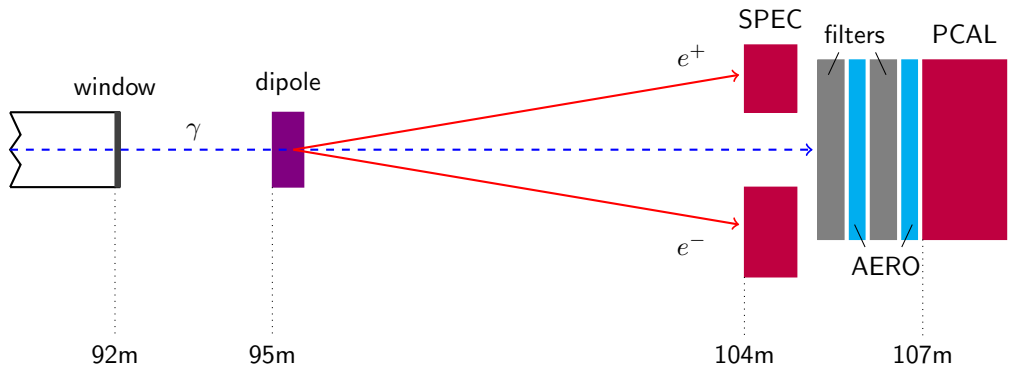}
\caption{The layout of the luminometers in the ZEUS experiment. 
The IP is on the left side in the
ZEUS detector. A dipole magnet just downstream of the beam-exit window deflected electrons 
originating from converted photons to the two electromagnetic calorimeters of the~\spec~SPEC. 
PCAL denotes the~\pcal~and filters the carbon absorber blocks. AERO are Cerenkov 
counters not used for the luminosity measurement.}
\label{fig-layout}
\end{figure}

Photons originating from bremsstrahlung at the IP moved through a vacuum chamber 
of 92 \met{} length
terminated by an exit window made
of an
aluminum alloy of 9.9~\mm{} thickness.
Approximately 9\,\% of the photons converted into 
electron pairs in this window.

Upstream of the window, inside 
the vacuum chamber, the aperture was restricted due to
beam-line components and magnets, which absorbed photons 
from the edges of the photon beam. The shape of the unobscured region was 
obtained from a foil sensitive to synchrotron radiation, which was located in 
the beam-line near the front face of the calorimeters~\cite{spec}.

\subsection{The Photon Calorimeter}
\label{sec-cal-design}
Photons not converted in the exit window hit the photon calorimeter,
a lead-scintillator sandwich calorimeter of 24 X$_0$ depth positioned
107~\met~downstream of the interaction region.
Two carbon 
blocks of 2 X$_0$ thickness each were placed in front of the calorimeter 
to absorb 
low energy synchrotron radiation photons.
The calorimeter was composed of 48 lead absorber plates of 
20$\times$20~\cm$^2$ size and 0.5  X$_0$
thickness interspersed with 2~\mm~thick scintillator sheets. 
Two wavelength shifter bars were attached at each side to the scintillator sheets.
They were equipped at the rear end by 
photomultipliers. 

Downstream of the 8-th absorber plate, where the shower maximum is expected, two planes 
of crossed scintillator fingers 
of 1~\cm~width and 3~\mm~thickness were positioned.
Each finger was read-out directly by a photomultiplier.
The energy resolution of the calorimeter was measured in an electron beam as 
14\%/$\sqrt E$ and 90\%/$\sqrt E$ without and  
with the 4 X$_0$ carbon absorber blocks in front, respectively.

\subsection{The Luminosity Spectrometer}
\label{sec-spec-design}

The electron pairs created in the exit window follow the original 
photon direction, until they encountered a 60~\cm~long  
dipole magnet with a field strength of 0.5 T 
where they were deflected from the photon beam direction. 
A few meters downstream they 
were registered by two calorimeters placed 
above (\up) and below (\dn) the {y-z} plane, 
with the \z{} axis pointing in the direction of the proton beam, the
\x{} axis towards the centre of HERA and the \y{} axis completing a right-handed
Cartesian system. 
The aperture restrictions upstream of the 
exit window prevent direct photon hits in the calorimeters.

Both calorimeters were composed in the~-\z~direction
of 3.5~\mm~thick tungsten absorber plates interspersed with scintillator 
layers of 2.6 \mm{} thickness.  
Each layer was 
divided into strips of 7.9~\mm{} width. The strips were 
alternately vertically and horizontally mounted, 
providing an \x{} and \y{} position measurement of a 
shower. The scintillator strips of each orientation 
were connected sequentially in the \z~direction
to one wavelength shifter bar equipped with a 
photomultiplier, forming a readout channel. 
Each calorimeter had 16 channels in \x{} direction. 
In \y{} direction, the \dn{} 
calorimeter had 15 channels and the \up{} calorimeter 11 
channels. The latter is due to the limited space between 
the photon beam and the proton beam pipe.
The calorimeters, both with a 
longitudinal depth of 24 X$_0$, 
were calibrated in an electron beam with energies between 1 and 6~\gev.
The energy resolution and calibration uncertainty were measured
as 17\%/$\sqrt E$ and 0.5\%, respectively.

\subsection{Data Acquisition and Data Storage}
\label{sec-spec-daq}
The data acquisition system, DAQ~\cite{bold_daq}, common for the \pcal{} and
the \spec{}, was independent 
from the ZEUS central DAQ. This allowed 
processing of large data samples, necessary for the 
luminosity measurement, without affecting 
the central DAQ performance.
Signals from each photomultiplier were digitised by 12--bit flash ADCs and transferred to 
four memory boards, MB, after each bunch crossing. 
To correct for pedestals, 
the ADC values stored from the previous bunch 
crossing were subtracted.

\subsubsection{The Photon Calorimeter Specific Part}
\label{sec-PCAL-daq}
The digitised signals from the two photomultipliers 
were
averaged and converted into energy using calibration constants 
obtained as described in 
section \ref{cal-meas}.
Rates of events were counted
on-line with energy thresholds of 10~\gev~and 17~\gev~
for pilot bunches\footnote{Pilot electron bunches 
have no corresponding proton bunches and hence can be used to measure background rates.} 
and colliding bunches.  In addition, total rates and rates for
colliding bunches were recorded for energy thresholds 
of 7.5~\gev, 12~\gev~and 16.5~\gev, respectively. These rates were used to
cross check the corrections applied for pile-up of more than one bremsstrahlung photon 
per bunch crossing
in the luminosity measurement.  
For each run a special record is written containing pedestals,  
signals from light test-pulses, on-line event rates, electron 
and proton currents in HERA and 
the active time of the ZEUS experiment. 
The signals from the scintillator fingers were digitised and recorded
to monitor the photon shower position.

\subsubsection{The Spectrometer Specific Part}

From the sum of photomultiplier 
signals
 of the up and down calorimeters the energies $E^{up}$ and
$E^{down}$ were calculated using calibration constants 
from special runs as described in section \ref{spec-meas-pe}.
Only events with $E^{up}$ and $E^{down}$ above 2~\gev~generated a trigger signal 
for the readout.

The MBs stored incoming data into a buffer to await the 
trigger signal. If the MB buffer was full, incoming data 
were discarded until buffer space was available, leading 
to dead-time of the system. Counters 
recorded the total number of HERA buckets and 
the number in which the MB buffer was active, allowing
a continuous dead-time 
correction for the luminosity calculation. To minimise dead-time, 
a trigger prescale factor 3 was applied.
The typical trigger rate at 
the given prescale settings was around 3\,\kHz.
The photon energy and the photon position were calculated and an 
event selection as described in section 
\ref{spec-meas} was performed. The accepted events were stored every  
16.5 seconds in an environmental record. 
For data quality monitoring, 10000  triggered events 
were saved per run. To create this 
sample, every 100-th event was selected, 
starting at the beginning of the run.

\section{Photon Measurements}
\label{sec-phot-measurements}

The~\pcal~measured the rate of unconverted photons
above certain energy thresholds, and the~\spec~the rate of converted photons 
when the conversion
electrons hit the calorimeters and trigger the readout.

\subsection{Measurement of the Photon Flux in the Photon Calorimeter}
\label{cal-meas}\

Photons were counted above the energy thresholds given in section
\ref{sec-PCAL-daq}. 
The relation between digitised and averaged photomultiplier signals in ADC counts
and the energy of the photon, $E_\gamma$, is parametrised as 
\begin{equation} \label{cal_calib}
ADC = C_{cal} \cdot E_\gamma \cdot (1 + f_{nl} (E_{max} - E_\gamma) \cdot E_\gamma + \Delta,
\end{equation} 
where $E_{max}$ is the maximum photon energy, equal to the electron beam energy, and 
$\Delta$ a pedestal offset due to synchrotron radiation.  
The calibration constant $C_{cal}$ and the parameter $f_{nl}$, accounting for deviations
from a linear response, were obtained from a fit 
of the bremsstrahlung energy spectrum predicted from Monte Carlo to data taken
in special runs 
without proton beam. Then only bremsstrahlung photons from interactions of the electron beam with the residual gas 
in the vacuum chamber contribute with relatively low rate to the spectrum and pile-up of several photons per bunch crossing is
negligible.
An example for a fit to the data distribution is shown in \Fig{~\ref{fig-beam-gas}.
Also shown is  
the Monte Carlo distribution.
\begin{figure}[!htb]
\centering
\includegraphics[width=.7\textwidth,height=7cm]{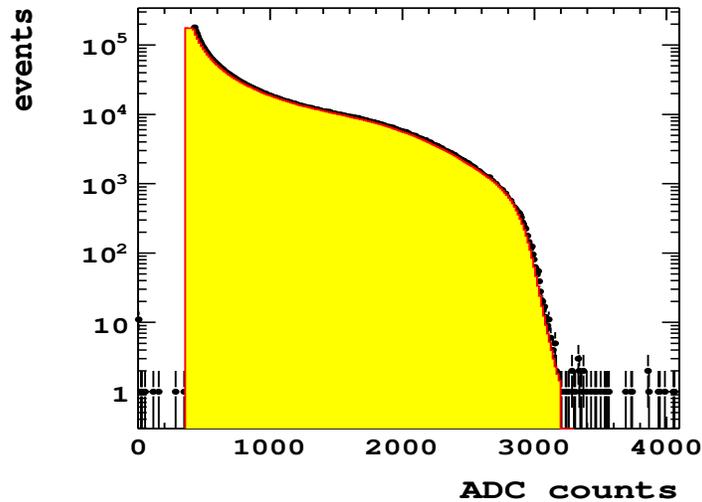}
\caption{The energy spectrum of photons in ADC counts obtained 
from interactions of the electron beam with residual gas.
Data (full black dots with error bars) are overlaid with the fitted 
Monte Carlo distribution (full red line on top of the yellow area). 
}                          
\label{fig-beam-gas}         
\end{figure}

During data taking with collisions, the photomultiplier gain was monitored with light pulses. Gain changes
were corrected by readjusting the high voltage.
In addition, using the rates of photons for different energy thresholds, the calibration constant was 
estimated to be stable within 1\,\% within a run.
 
The parameter $f_{nl}$ was determined to be -0.017 and found to be stable within all data taking periods. 
The parameter $\Delta$ was measured using random triggers as a function of the electron beam 
current per bunch crossing. 

Taking the event rates recorded for each of the energy thresholds and the 
electron pilot bunches,
the rate of bremsstrahlung events, $R_{\gamma}$, corrected for 
interactions of the beam with residual gas, 
was obtained as
\begin{equation} \label{cal_meas1}
R_{\gamma} = R_{tot}-R_{pilot}\cdot \frac{I_{e}^{tot}}{I_{e}^{pil}},
\end{equation} 
where $R_{tot}$ and $R_{pilot}$ were the total photon rate and the rate from the pilot bunches,
and $I_{e}^{tot}$ and $I_{e}^{pil}$ the total and the pilot-bunch electron currents, 
respectively. 

At high photon rates more than one photon from a single bunch crossing may have hit the calorimeter. 
The average 
number of photons per bunch crossing, $\bar n_\gamma$, was estimated to be
\begin{equation} \label{cal_meas2}
\bar n_\gamma = \frac{R_{\gamma}\cdot 
\sigma_b }{\sigma(E_\gamma > E_{tres.})\cdot 
f_r \cdot N_{cb}},
\end{equation} 
where $ \sigma_b$ and $\sigma(E_\gamma > E_{tres.})$
are the total bremsstrahlung cross section\footnote{$ \sigma_b$ was calculated for photons with energies above
0.1 \gev, being the energy threshold of the calorimeter} and the
bremsstrahlung cross section above 
an energy threshold $E_{tres.}$, $f_r$ is the HERA orbit frequency
and $N_{cb}$ the number of colliding bunches. 
The bremsstrahlung rate corrected for multiple photons per bunch crossing was then
\begin{equation} \label{cal_meas3}
R'_{\gamma} = R_{\gamma}\cdot (1+P(E_{tres.},\bar n_\gamma)),
\end{equation}  
where the correction term {\it P}($E_{tres.},\bar n_\gamma$) was determined from Monte Carlo simulations.

\subsection{Measurement of the Photon Flux in the Spectrometer}
\label{spec-meas}
\subsubsection{Position and Energy Measurement of Electrons and Photons in the Spectrometer}
\label{spec-meas-pe}

At nominal magnetic field,  
electrons with energies 
above 3.5~\gev~hit the calorimeters.
Events with electrons measured in both calorimeters were 
used to reconstruct the energy and 
transverse coordinates of the converted photon.

To minimise the impact of noise, only 4 strips near the shower 
maximum were used for the electron energy measurement. 
The measurement was corrected for 
transverse energy leakage, a small effect except 
at the edges of the calorimeter, and the light 
attenuation in scintillators. The photon energy was calculated as the sum of 
electron energies.

The shower position was calculated as 
the energy-weighted average in \x~and \y. Only channels 
with an energy above  
60~\mev{} were used. The photon coordinates,
$X_{\gamma}$ and $Y_{\gamma}$,
were obtained from the position and energy measurements 
of the individual electrons, taking into account 
electron deflection by the dipole magnet.

The calorimeters were routinely calibrated to take into account 
gain variations due to radiation damage and 
recovery mechanisms in the scintillators and wavelength shifter bars. 
Special runs were taken with an adjustable slit collimator 
inserted into the photon beam to select a certain \y{} position. Using the bending power of the
dipole magnet a direct relation between
the electron energy and the \y{} position of the shower in the calorimeter 
was obtained.
Data taken in this configuration were used to 
determine the individual channel gains iteratively. 
After each step gains were varied to match the predicted energy and
\y{} position. 
The procedure 
converged after a small number of steps.

\subsubsection{Event selection     }
\label{sec-spec-evnt}                                              
The depositions in the spectrometer of a typical 
triggered event are shown in \Fig{~\ref{fig-event}}.
After the reconstruction of the energy
and position of the showers in 
each calorimeter the following 
selection criteria were imposed:
\begin{itemize}
\item The reconstructed energy for each calorimeter 
must be larger than 3.5~\gev{} to suppress noise.
\item The largest energy depositions must not be 
at the edge strips in each of the four planes to 
minimise leakage, thus ensuring good energy 
and position reconstruction.
\item The \y{} 
position of a shower in the \dn{} detector was restricted 
to make the fiducial regions of \up{} and \dn{} calorimeters equal.
\item The rms width of the shower must be less 
than 1~\cm{} for each of the four projections (\rmscut) 
to reject hadron showers from proton interactions with the residual gas.
\end{itemize}
\begin{figure}[!htb]
\centering
\includegraphics[width=\textwidth]{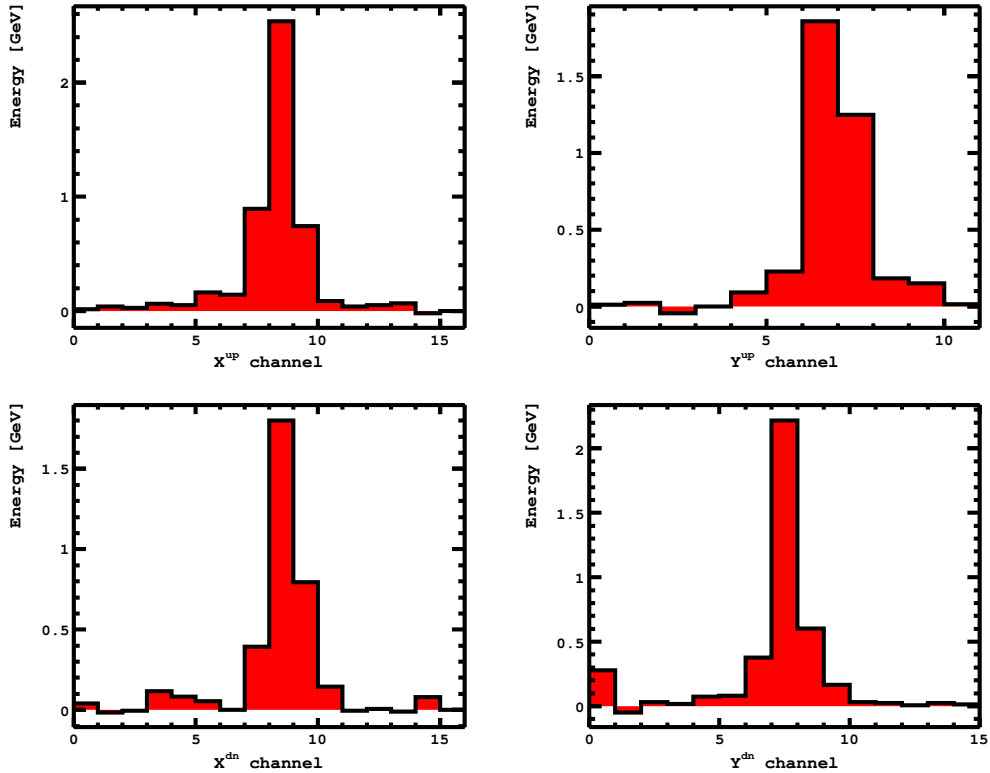}
\caption{A spectrometer event with an electron pair 
coincidence. The energy deposition per channel 
in the \x{} (left) and \y{} (right) directions
for the \up{} (top) and \dn{} (bottom) calorimeters. 
Negative energy in several channels result from 
the subtraction of the stored ADC values of the 
previous bunch crossing.}
\label{fig-event}
\end{figure}
The last requirement was applied only in the data taking periods 2005 and 2006 with electron beam.
It 
caused severe event loss
due to depositions from synchrotron radiation at the inner edges of the \y{}-plane.
This loss 
of events 
was corrected for using the 10000 event sample recorded 
at the beginning of each run~\cite{drugakov_these}.  
In data taking periods with positron beams in 2006 and 2007 the last requirement was omitted, since
the fraction of hadron showers in the spectrometer was estimated to be negligible
using runs with only protons in the collider.

About half of the triggered events passed the 
requirements.
The reconstructed photon coordinates, $X_{\gamma}$ and $Y_{\gamma}$ 
and the photon energy were stored into histograms
for each environmental record. The mean photon beam position and width
were calculated at the exit window.
From the moments of the $X_{\gamma}$ and $Y_{\gamma}$ distributions 
and the correlation coefficient $<$$X_{\gamma}$Y$_{\gamma}>$ the photon beam ellipse and its tilt 
were determined.
The number of photons was counted 
and corrected for interactions of the beam with the residual gas using \Eq{(\ref{cal_meas1})}.
The raw data of the environmental record 
were then discarded.  

\section{Monte Carlo Simulations to Determine the Acceptances}
\label{monte-carlo}\

The acceptances of the~\pcal~and~\spec~were
determined by Monte Carlo simulations  
using the GEANT~3.21 package.
Photons were generated according to the bremsstrahlung 
spectrum as given in \Eq{(\ref{eq-bh})} taking into account the 
electron beam angular divergence. They were then
randomly distributed in a sufficiently large area of the (x,y) plane
at the z-position of the exit window.  
Using 
beam profiles 
measured by the spectrometer as described in
section~\ref{sec-spec-evnt} 
and the aperture limits of the vacuum chamber,  
a weight was applied to each photon. 
Only photons generated 
inside the measured aperture were tracked to the luminometers.

\subsection{Photon Calorimeter Simulation}

Photons with an energy larger than $E_{min}$ = 0.1~\gev~not converted in the beam exit window or air
were transported to the photon calorimeter and a shower was simulated.
The energy deposited in the scintillators and scintillator fingers was recorded
for shower energy and position reconstruction.
The light collection and the impact of the readout electronics 
was parametrised using test-beam
data. 

A detector response function $F(E_{cal})$ was calculated as a function of the 
photon energy $E_{\gamma}$,
\begin{equation}
 F(E_{cal}) = 1/\sigma_{b}\times\int_{E_{min}}^{E_{e}-m_{e}}\frac{d\sigma}{dE_{\gamma}}P(E_{\gamma},E_{cal})dE_{\gamma},
\end{equation}  
with $P(E_{\gamma},E_{cal})$ being the probability that for an incident photon with an energy
$E_{\gamma}$ the energy $E_{cal}$ was measured in the calorimeter and $\sigma_{b}$ the 
bremsstrahlung cross section integrated from $E_{min}$ to the upper kinematic limit.
Using the detector response function the cross section, $\sigma_{thr}$,
for photons with a measured 
energy above a threshold, $E_{thr}$, is obtained, 
\begin{equation}
\sigma_{thr}(E_{thr}) = \sigma_{b}\int_{E_{thr}}^{\infty}F(E_{cal})dE_{cal}.
\end{equation}  
The acceptance $A$ is the number of events above a given energy threshold, $N_{thr}$, divided by the 
generated number of events, $N_{gen}$,
\begin{equation}
A = \frac{N_{thr}}{N_{gen}}= \int_{E_{thr}}^{\infty}F(E_{cal})dE_{cal}.
\end{equation} 

\subsection{Spectrometer Simulation}
\label{sec-spec-mc}

The generated photons were converted in the exit window. 
The electrons were transported to the calorimeters using a parametrised 
field map of the dipole and taking into account
interactions 
in the exit window and in air.  
Showers were simulated in the
calorimeters and the energy depositions in the scintillator tiles were transformed into 
signals of the photomultipliers.
The
simulated events were subject to the reconstruction and selection 
as described in section 
\ref{spec-meas} and accepted events were used to
to calculate  $X_{\gamma}$ and $Y_{\gamma}$, and the 
energy of the photon, $E_{\gamma}$. 
This sample 
was then used to calculate the acceptance for every \record{}. 
Assuming a Gaussian shaped photon beam and taking into 
account the correlations between 
$X_{\gamma}$ and $Y_{\gamma}$ determined from data, 
the   
events were re-weighted to fit to the measured $X_{\gamma}$ and $Y_{\gamma}$ 
distribution. 
An example of the result is shown in \Fig{~\ref{fig-beamprof}}.
The acceptance was then calculated as the ratio of the sum of accepted and 
generated Monte Carlo event weights. A typical value 
of the average acceptance is 0.65\% for $E_{\gamma}>8$~\gev~with a maximum 
of 2\% at 22~\gev.

\section{Luminosity Measurement}

\subsection{Calorimeter Procedure}

The counter rates with the energy thresholds 
defined in section~\ref{sec-PCAL-daq}
were 
corrected for interaction of the electron beam with residual gas in the vacuum and pile-up using \Eq{~(\ref{cal_meas1})} and
\Eq{~(\ref{cal_meas3})} and summed up to $N_\gamma$ as defined in \Eq{~(\ref{eq-lumi})}. 
For each energy threshold a value for the luminosity was obtained.
Finally a correction for the ZEUS active time was applied.

\subsection{Spectrometer Procedure }
\label{sec-specanalysis}

\begin{figure}[!htb]
\begin{tabular}{cc}
\includegraphics[width=0.47\textwidth]{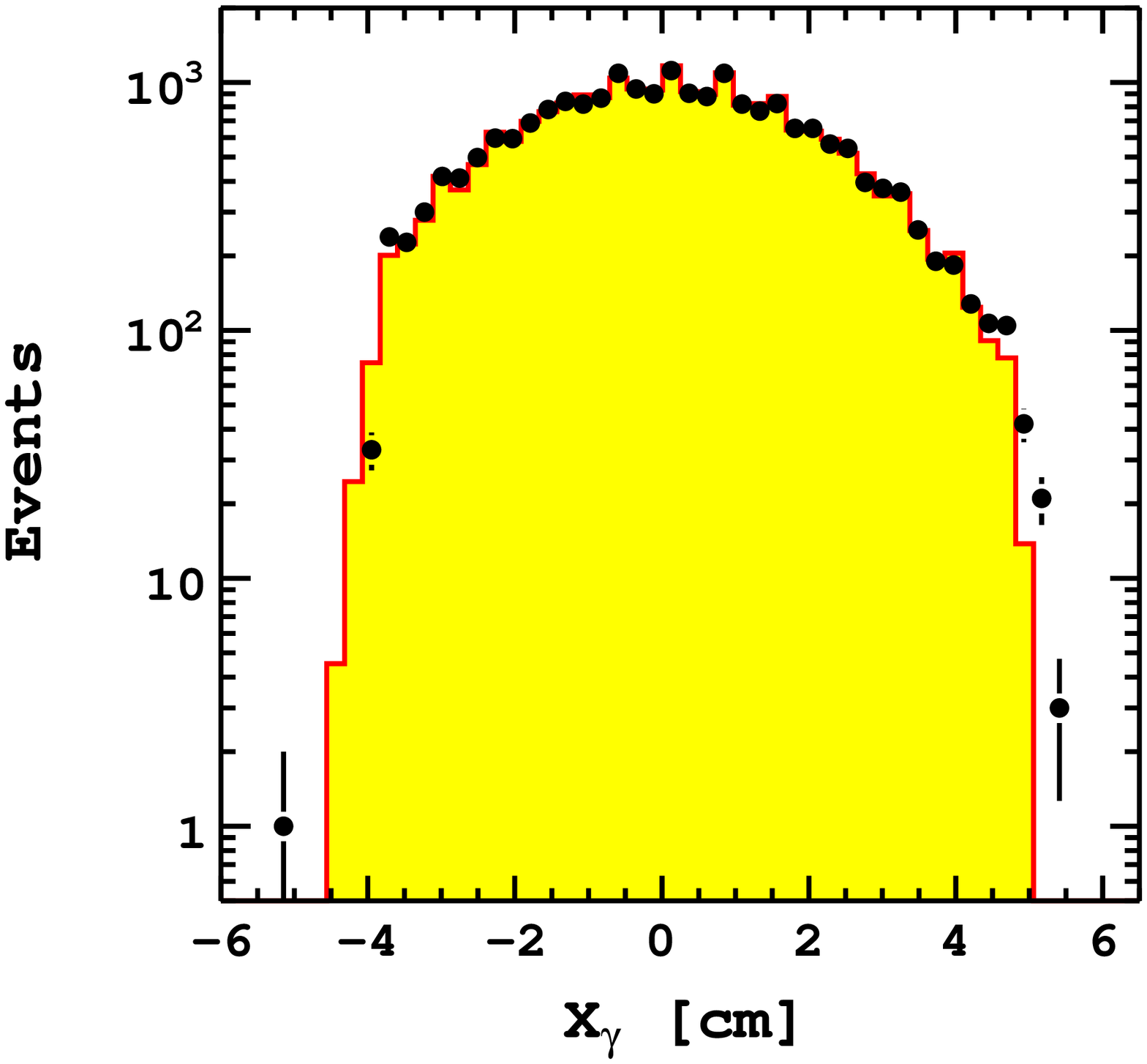}&
\includegraphics[width=0.47\textwidth]{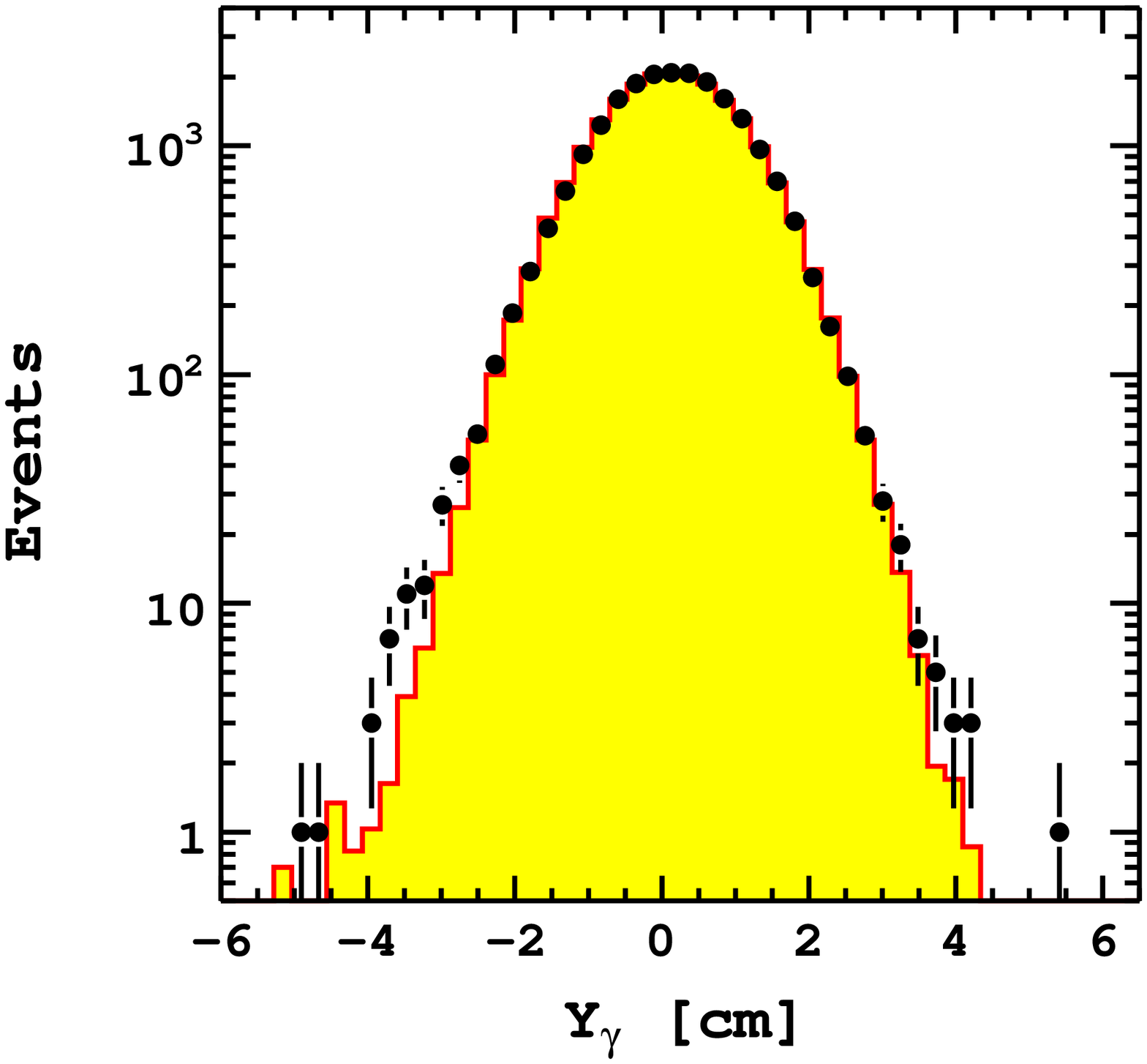}
\end{tabular}
\caption{The reconstructed $\rm{x}$ (left) and $\rm{y}$-positions (right)
of accepted photons 
in the~\spec~ collected 
in one \record{}. Dots with error bars are data and 
the histogram is the Monte Carlo prediction.}
\label{fig-beamprof}
\end{figure}
The number of  
photons was counted in each \record{} applying the criteria given in  
section~\ref{sec-spec-evnt}.
The integrated luminosity was calculated using \Eq{~(\ref{eq-lumi})} 
and corrected for dead times of the ZEUS and the spectrometer DAQ systems. 
The impact of pile-up due to conversion of more than one photon 
per bunch crossing was
estimated as described in the Appendix. 
The number of photons obtained 
after applying the selection criteria is slightly larger than the true number. 
A correction, as
shown in~\Fig{~\ref{fig-pileup}}
as a function of a quantity $r$, is applied, where $r$ is 
the fraction of colliding bunches with 
a bremsstrahlung event passing the selection criteria.
The correction, reaching 0.4\,\% at $r$ = 0.0015, 
corresponding to the maximal instantaneous luminosity measured
at HERA,
\begin{figure}[!tbh]
\centering
\includegraphics[width=0.47\textwidth]{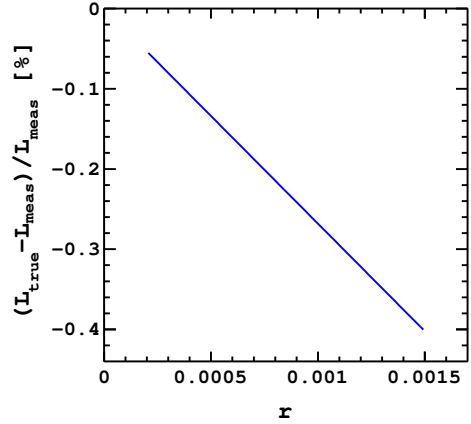}
\caption{Luminosity shift due to pile-up as a function of 
the fraction of colliding bunches with a bremsstrahlung event
passing the selection criteria. The 
shown range of $r$ corresponds to the experimental 
situation. The simulation was performed for the photon 
beam with average parameters.}
\label{fig-pileup}
\end{figure}
was applied to the 
measured luminosity of each \record{}.

\section{Systematic Effects in the Luminosity Measurement}
\label{sec-revi_recon}

\subsection{Systematics Common to the Photon Calorimeter and the Spectrometer}

\subsubsection{Acceptance}

The estimation of the uncertainty in the acceptance holds for both the~\pcal~and the~\spec.  
The shape of the aperture area and its alignment relative to the detectors were each 
measured to an accuracy of 1~\mm. A variation of these quantities by 1~\mm~  
shifted the values of the acceptance  
within 1\%.

The variation of the acceptance as a function of the photon beam position
was studied using neutral current events
selected in the ZEUS detector. 
For the 2004/06 electron beam and the 2006/07 positron beam 
independent analyses were done using samples of neutral current events with 
Q$^2$ $>$ 185~\gev$^2$.
This process has a high rate and is not expected to have any common systematics with 
the luminosity measurement.
Details of the event selection criteria can be found in 
Ref.\cite{nc-event}.

The ratio between the rate of neutral current events and the luminosity measured with the 
spectrometer is shown in \Fig{~\ref{fig-NLx}} (left) as a function of the x position 
of the photon beam, $X_\gamma$,
and in \Fig{~\ref{fig-NLy}} (left) as a function of the y position of the photon beam, $Y_\gamma$
for data taken with a positron beam.
\begin{figure}[!htb]
\begin{tabular}{cc}
\includegraphics[width=0.47\textwidth]{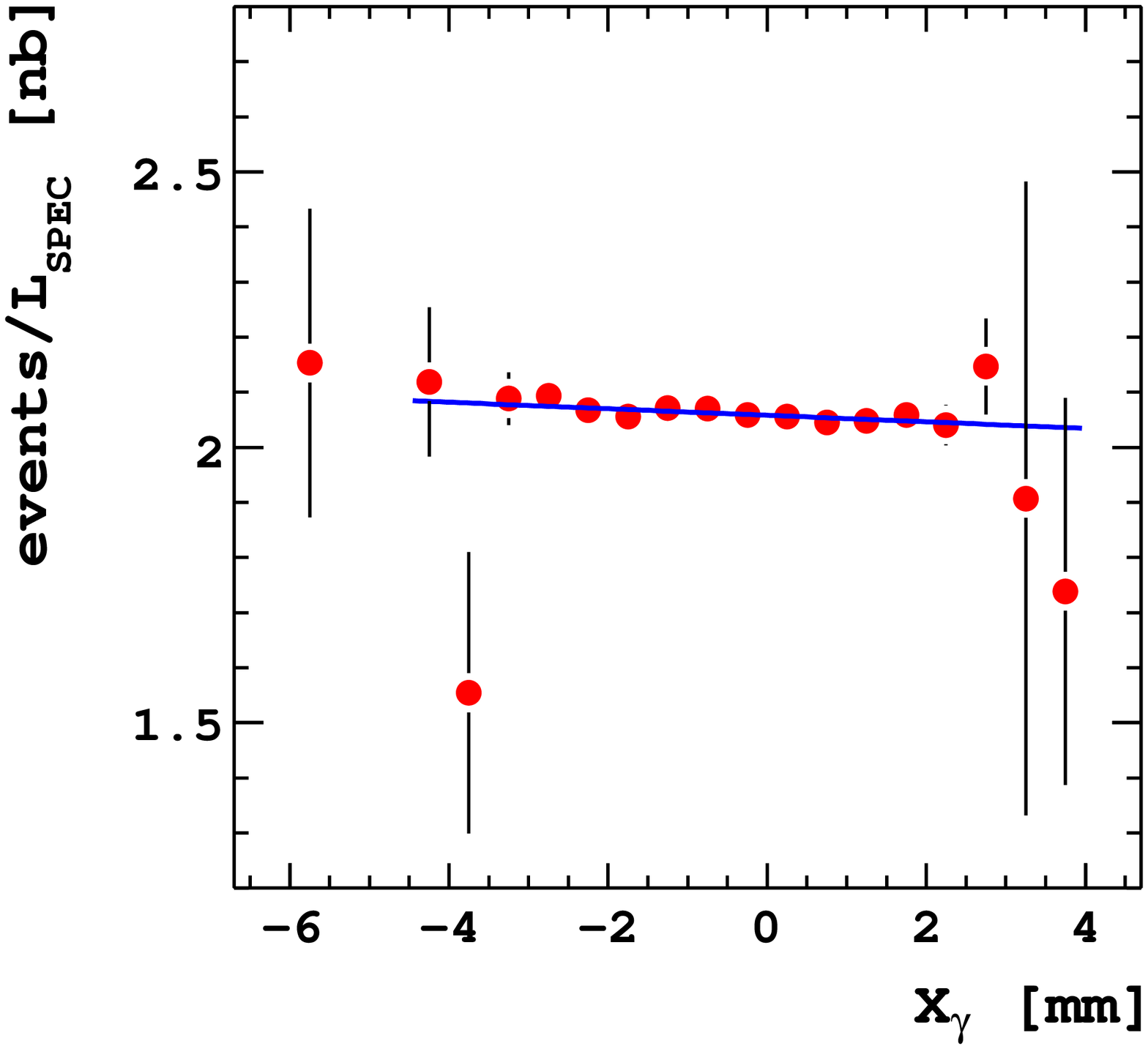}&
\includegraphics[width=0.47\textwidth]{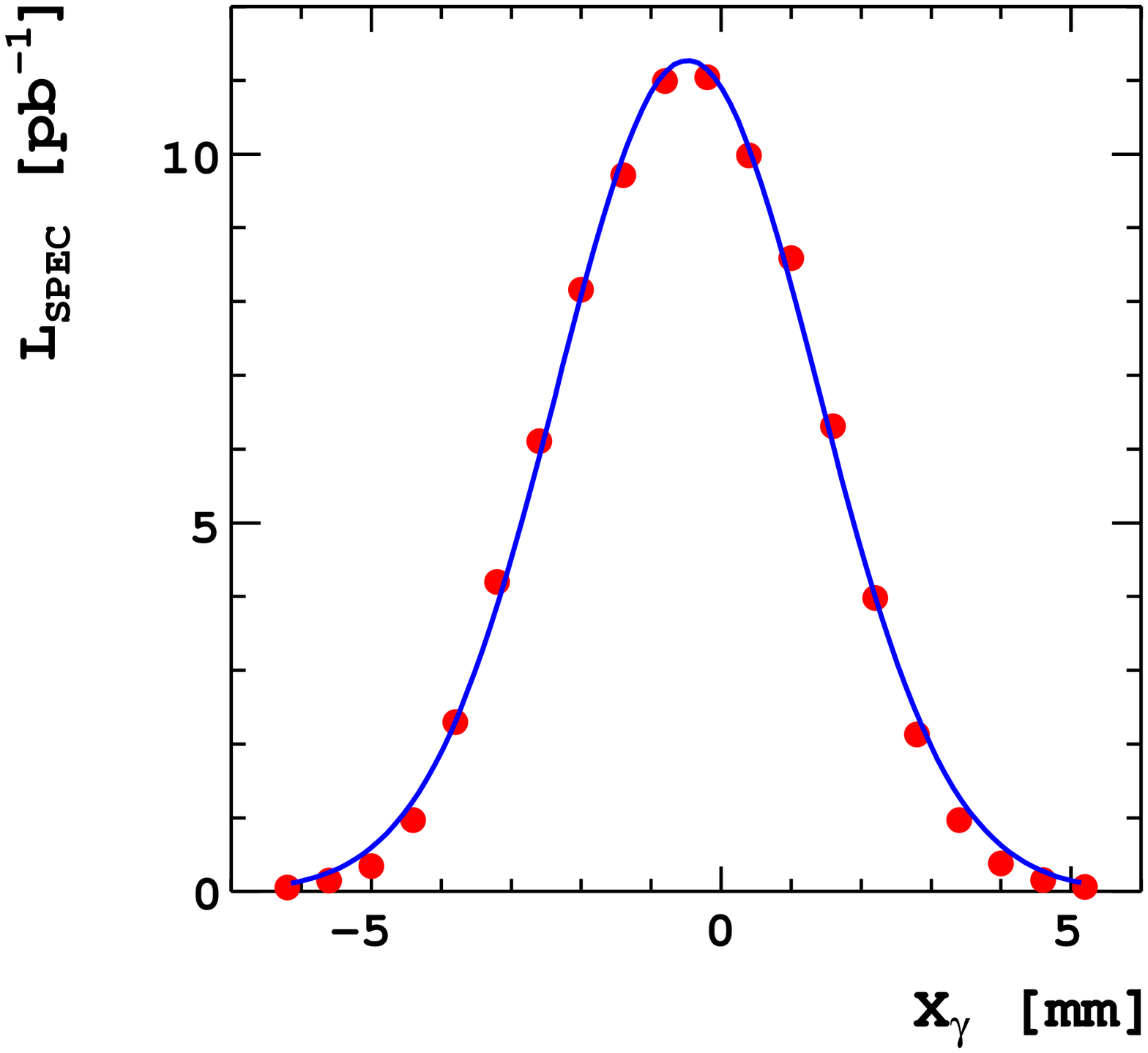}
\end{tabular}
\caption{Left: The ratio of the neutral current event rate
to the luminosity measured in the spectrometer as a function 
of the \x{} position of the photon beam. 
The distribution is fit with a polynomial of 
degree one. Right: The distribution of photons weighted with the luminosity
as a function of the \x{} position of the photon beam. The distribution is fit with a Gaussian.}
\label{fig-NLx}
\end{figure}
\begin{figure}[!htb]
\begin{tabular}{cc}
\includegraphics[width=0.47\textwidth]{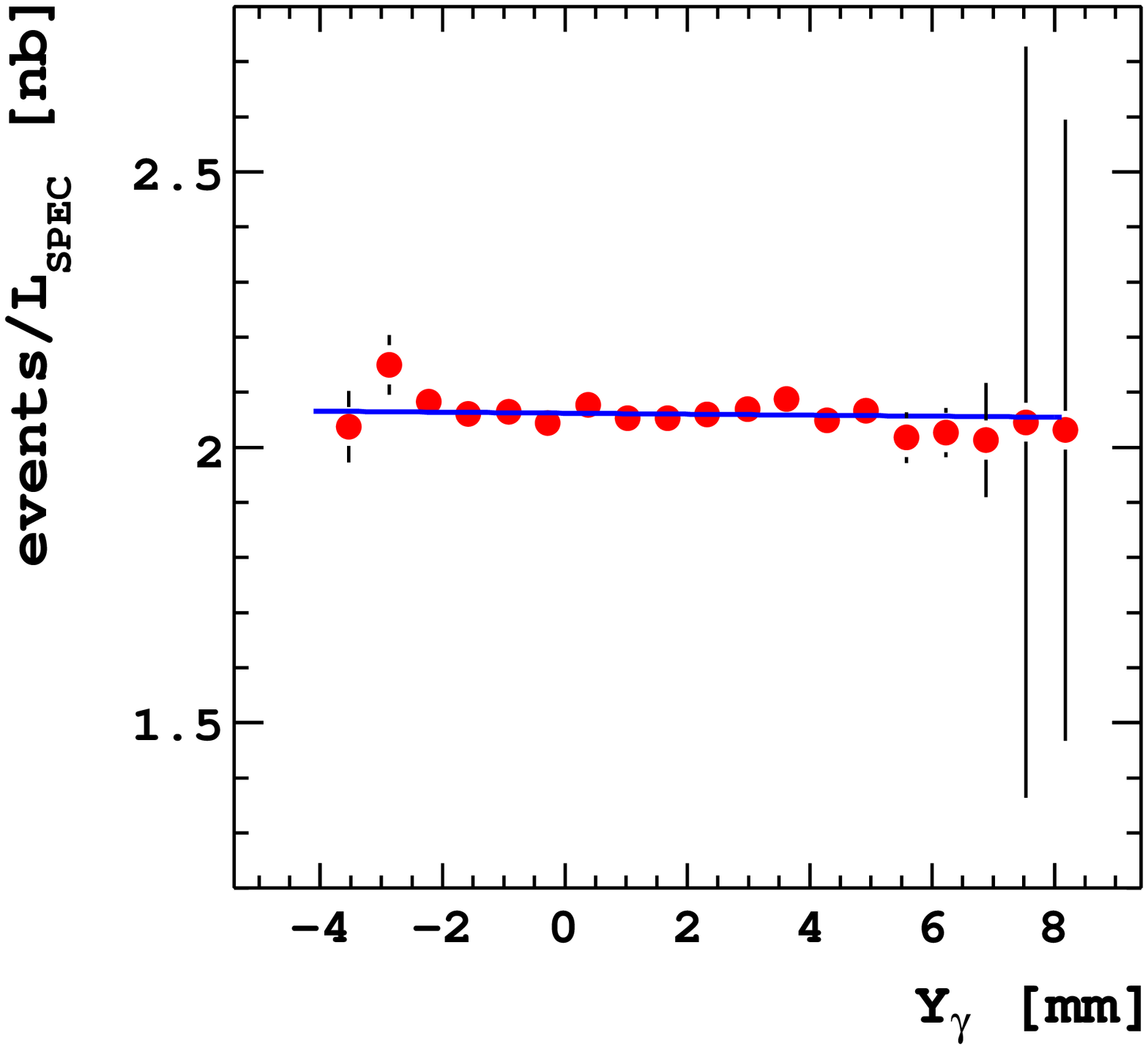}&
\includegraphics[width=0.47\textwidth]{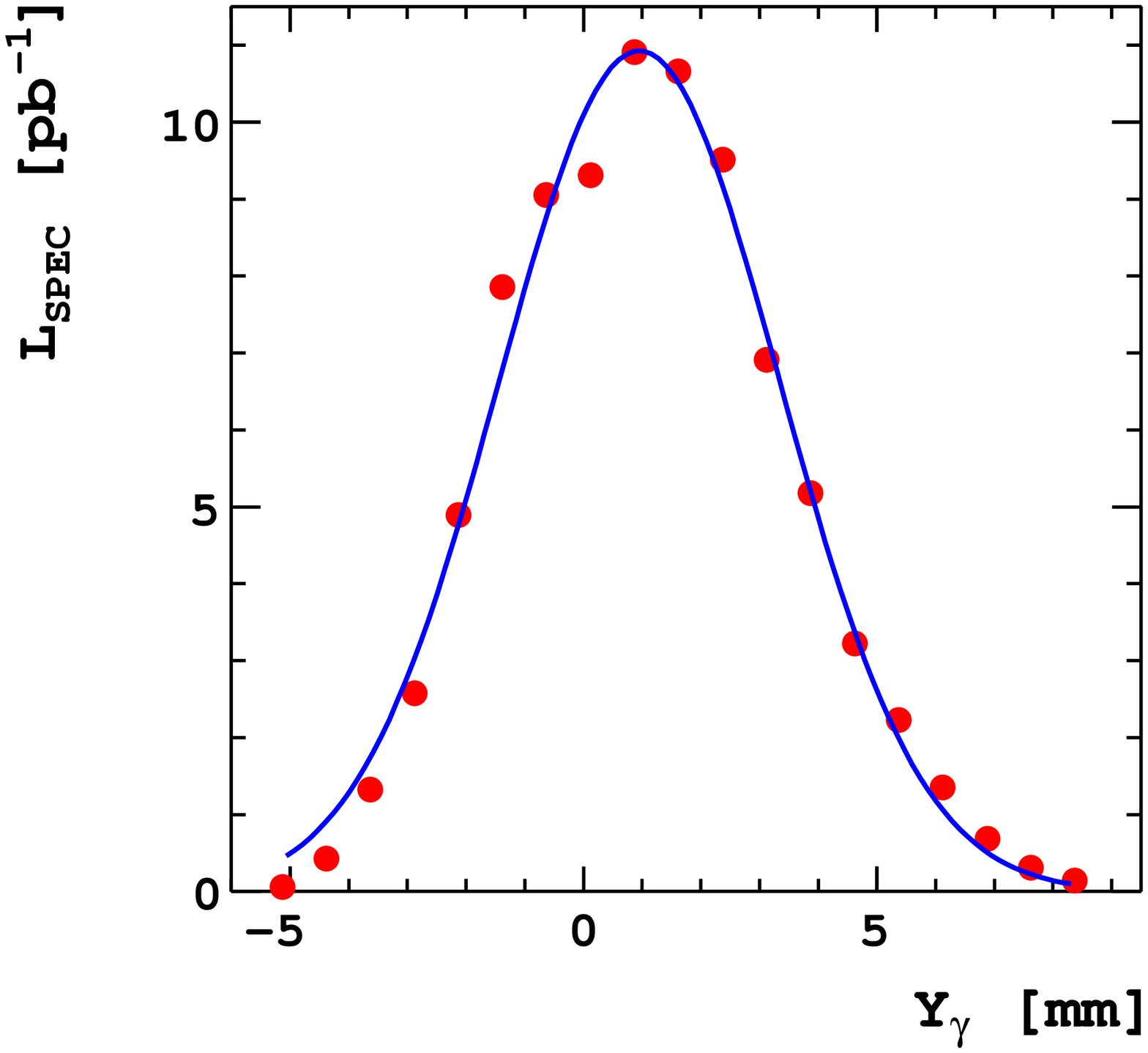}
\end{tabular}
\caption{Left: The ratio of the neutral current event rate
to the luminosity measured in the spectrometer as a function
of the \y{} position of the photon beam. 
The distribution is fit with polynomial of degree one. Right: 
The distribution of photons weighted with the luminosity as a function of the \y{} 
position of the photon beam. The distribution is fit with a
Gaussian.}
\label{fig-NLy}
\end{figure}
The ratio as a function of $X_\gamma$ exhibits a slope of $(-5.9 \pm 4.2) \times 10^{-2}\rm{cm}^{-1}$
whereas no dependence on $Y_\gamma$ is observed. 
  
Also shown in the right figures is the distribution of photons used in the luminosity measurement 
as a function of $X_\gamma$ and $Y_\gamma$,
normalised to the integrated luminosity. 
A fit with a Gaussian to the distribution shown in \Fig{~\ref{fig-NLx}} (right) results in a mean value of 
-0.5~\mm~and a width
of 1.9~\mm. The systematic uncertainty of the luminosity is estimated as the variation within the width
as 1.1\%.  

The same study using data taken with the electron beam results in a systematic uncertainty of 1.2\%.

The impact of the event rate on the luminosity measurement
was investigated using the ratio between neutral current events and the luminosity 
measured by the~\spec~as a function of the product of the electron and proton currents. 
These currents are measured 
with other devices, and their product is proportional to the luminosity.   
Using the same technique as described above a systematic uncertainty of 0.6\% was estimated for 
data taken with positron beam. No impact was observed for data taken with electron beam.

These results obtained using the~\spec~were considered to be valid also for the~\pcal.

\subsubsection{Theory Uncertainty}
\label{sec-theory}

The formula given in \Eq(\ref{eq-bh}) for the bremsstrahlung cross
section as a function of the photon energy is 
obtained in the Born approximation neglecting 
the finite size of the proton. Recently the process has been calculated including 
one-loop QED radiative corrections and proton form-factors~\cite{makarenko}.
The inclusion of the one-loop QED corrections causes a small reduction in the photon energy 
spectrum at the high energy limit of the acceptance. 
The angular spread of the photon momenta obtained from 
this calculation is by an order of magnitude
smaller than the divergence of the electron beam
in ZEUS. The impact of both effects on the luminosity measurement is negligible. 

\subsubsection{Summary of Common Systematic Uncertainties}
A summary of the systematic uncertainties common to the Photon Calorimeter and the Spectrometer is given in Table~\ref{tab:common}.

\begin{table}[h]
    \caption{The systematics uncertainties of the luminosity measurement common to the  
    photon calorimeter and the spectrometer 
    for  electron and positron beams. All values in \%.}  
    \label{tab:common}
    \begin{center}
      \begin{tabular}{|l|c|c|} \hline
        Source of systematics &  2005/2006 $e^-p$    &   2006/2007 $e^+p$   \\ \hline
        aperture and          &  1                   &     1                \\
        detector alignment    &                      &                      \\
        x-position of the     &   1.2                &   1.1                \\
        photon beam           &                      &                      \\
        beam currents         &    -                 &   0.6                \\ \hline                         
        sum                   &   1.6                &  1.6                 \\ \hline
      \end{tabular}
    \end{center}
  \end{table}

\subsection{Photon Calorimeter}

The systematic uncertainties originating from the calibration of the \pcal~and 
multiple photon hits were considered. 

\subsubsection{Pedestal Shifts and Calibration Constants}
In the calibration of the \pcal~an offset $\Delta$ due to synchrotron radiation in the photomultiplier signal 
was taken into account. 
The synchrotron radiation was measured with random triggers of bunch crossings and parametrised
as a function of the electron beam current. Values for pedestal shifts of 
0.20 \mev/$\mu A$ and 0.08 \mev/$\mu A$ 
were obtained for electron and positron beams, respectively.
The impact of a remaining pedestal uncertainty of 85~MeV and the uncertainty on the calibration constant
was translated in an uncertainty of the luminosity measurement.
The combination of both effects lead to
a systematic uncertainty of 1.5\%.

\subsubsection{Multiple Photon Hits}

The luminosities obtained for several energy thresholds
were corrected for pile-up using 
\Eq(\ref{cal_meas3}).
The differences of the values obtained for several energy thresholds 
were distributed as a function of the       
average number of photons per bunch crossing. 
From their variation a systematic uncertainty of 0.5\% was estimated.

\subsection{Spectrometer}
\label{sec-spec-sys}

As additional sources of systematic uncertainties specific for the~\spec~
the
proton beam interactions with residual gas in the beam-pipe, potentially 
creating depositions in the calorimeters, and the
conversion probability in the beam exit window were considered.

\subsubsection{Impact of Proton Beam Interactions with Residual Gas in the Beam-pipe}

Data 
was collected when only protons were circulating 
in HERA
to estimate the contribution of background 
from interactions of the proton beam with residual gas in the beam-pipe.  
The event selection criteria, described in 
\sect~\ref{sec-spec-evnt}}, were applied. The event rate without 
applying the \rmscut{} was 0.1 \Hz. 
The typical event rate for bremsstrahlung events 
was 10-15 \kHz{} at the beginning of a 
fill. Hence the fraction of events from proton--residual gas
interactions 
in the data is about $10^{-5}$. 

The effect on the 
luminosity measurement 
due to pile-up of a proton shower 
in the \up{} calorimeter and an electron shower in 
the \dn{} calorimeter
was estimated from the data to be 
less than $10^{-6}$ at standard vacuum conditions.

\subsubsection{Photon Conversion Coefficient in the Beam Exit Window}
\label{sec-exit-window}

The beam exit window, where about 9\% of the photons were expected to convert, 
consists of an aluminum alloy. After the shutdown of HERA the window
has been dismounted and its thickness and composition
measured. 
For the thickness 9.9 $\pm$ 0.01~\mm~was obtained.
The chemical composition of the beam exit window was determined by two laboratories
using mass spectroscopy.
The measurements
are summarised in Table~\ref{tab1}.
\begin{table}[hbt]
\begin{center}
\caption{The chemical composition of the beam exit window measured in two 
laboratories and the 
photon conversion coefficient
$\mu_{pair}$ for photons of 20 \gev~in the elements contained. 
} 
\label{tab1}
\vspace{0.1cm}
\begin{tabular}{|l | c | c | c | c |}
\hline
Element      & Z          &  $\mu_{pair}$ cm$^{-1}$ & fraction, \% (lab1)& fraction, \% (lab2) \\       
\hline   
Mg           & 12         &  0.0528     &  0.318            & 0.344       \\
Al           & 13         &  0.0858     &  81.790           & 86.200       \\
Si           & 14         &  0.0815     &  14.765           & 10.300       \\
Fe           & 26         &  0.4385     &  0.227            & 0.260       \\
Cu           & 29         &  0.5342     &  2.433            & 2.400        \\
Zn           & 30         &  0.4397     &  0.338            & 0.365        \\  \hline
average $\mu   $ &        &             &  0.0984           & 0.0986       \\
\hline
\end{tabular}
\end{center}
\end{table}
Although there are differences of up to several per cent in the chemical 
compositions measured in two
laboratories,
the average photon conversion coefficients at 20 \gev~are equal within 2.3 per mille.
In the following the measurements of lab2 are used. 
\begin{figure}[!tbh]
\centering
\includegraphics[width=0.47\textwidth]{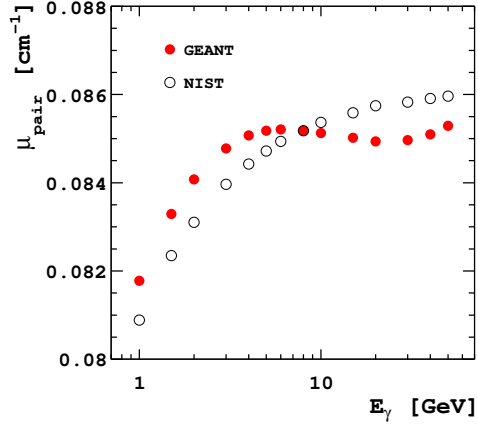}
\caption{The photon conversion coefficient of aluminum as a function of the photon energy.
The full red dots are obtained from the parametrisation in GEANT4 
and the open dots from tables published by NIST.
}
\label{fig-conversion}
\end{figure}
The photon conversion coefficient is a function of the energy. 
In~\Fig{~\ref{fig-conversion} its parametrisation 
used in GEANT{\footnote{GEANT3 and GEANT4 use the same cross section parametrisation.} 
is compared
with the values from the NIST~\cite{NIST} data base
for aluminum being the largest element fraction in the exit window. In the relevant
energy range around 20 \gev~the difference in the conversion coefficient approaches 1\%.

For the acceptance calculation the values from the NIST database are used.  
The uncertainty of the luminosity resulting from the photon conversion coefficient 
in the beam exit window is estimated to be 0.7\%. The dominant contribution originates from the 
uncertainty of the cross section measured at 1.5~\gev~\cite{c_s_nist} used in the NIST database and its extrapolation to 
20 \gev.
The uncertainty of the 
thickness contributes with 0.1\% and the uncertainty of the chemical composition with 0.3\%.

\subsection{Summary of the Systematic Uncertainties}

The systematic uncertainties estimated for the \pcal~were the same 
for electron and positron beams.
The uncertainty not common with the uncertainties of the 
\spec~results to 1.6\%.  

The systematic uncertainties obtained for the spectrometer are estimated for
electron and positron beams separately.
The uncertainty not common with the  \pcal~amounts to 0.9\%.
A summary is given in Table \ref{tab:system}.

\begin{table}[h]
    \caption{The systematics uncertainties estimated for the luminosity measurement in the
    photon calorimeter and the~\spec~for
    electron and positron beams. All values in \%.}  
    \label{tab:system}
    \begin{center}
      \begin{tabular}{|l|c|c|c|} \hline
        Source of systematics & photon   & spectrometer        &   spectrometer   \\
                              & calorimeter& 2005/2006 $e^-p$    &   2006/2007 $e^+p$  \\  \hline
        common systematics     &    1.6 & 1.6      &  1.6                \\
                               &         &         &                      \\
        photon conversion     &          & 0.7       &    0.7               \\
        in the beam exit window &        &           &                      \\
        rms-cut correction    &          & 0.5       &     -                 \\
        beam currents         &    -     &           &    0.6                \\ 
        pedestal shifts       &   1.5    &           &                       \\
        pile-up               &   0.5    &           &                      \\ \hline
        sum                   &   2.2    &   1.8     &  1.8                 \\ \hline
      \end{tabular}
    \end{center}
  \end{table}

Averaging the systematic uncertainties not common between \pcal~and \spec~ 
and adding it in quadrature to the common systematic uncertainty leads to a 
total systematic uncertainty of the luminosity measurement
of 1.7\%.

\section{Comparison of the Luminosities Measured with the Calorimeter and Spectrometer}

The distribution of the ratio between 
the luminosity values obtained for each run from the~\spec~and the 
\pcal~is shown in \Fig{~\ref{fig-LLe-}} (left) 
for data taken with electron beam 
and in \Fig{~\ref{fig-LLe+}} (left) for data taken with positron beam.  \\
Very good  
agreement between the two measurements is found.
The distributions
of the ratios have nearly Gaussian shape with a mean value 
of 1.01 both for the electron 
and positron data.

The distributions of the ratio versus run number is nearly flat over the 
periods as shown in \Fig{~\ref{fig-LLe-}}~and \Fig{~\ref{fig-LLe+}}~(right), 
indicating a very good stability of 
the measurements. Most of the points lie within 1\,\% of the 
central values, with a slightly larger spread for runs around 54000
in which corrections are applied for event losses due to 
the \rmscut~used in the event selection for the~\spec.
\begin{figure}[!htb]
\begin{tabular}{cc}
\includegraphics[width=0.47\textwidth]{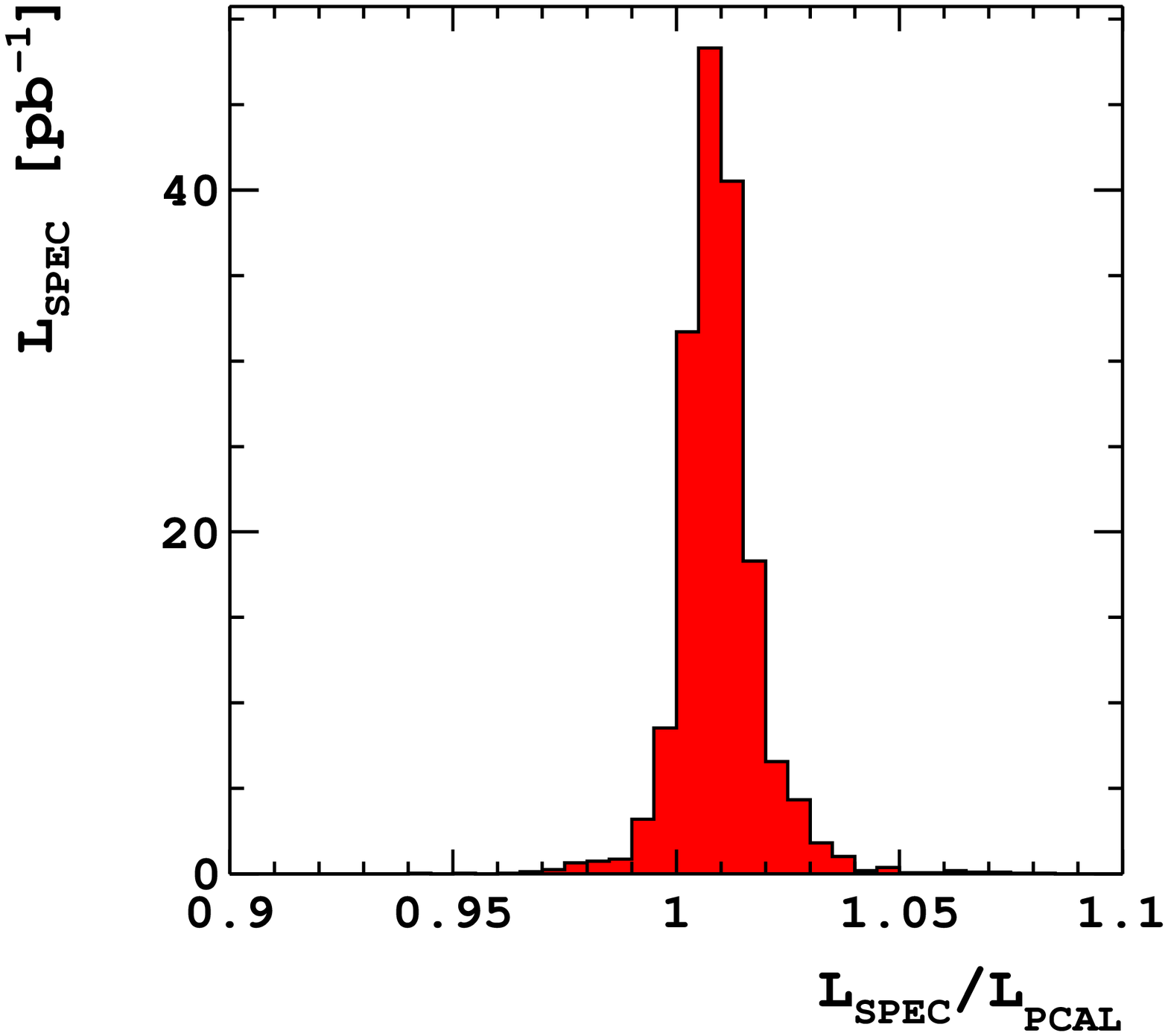}&
\includegraphics[width=0.47\textwidth]{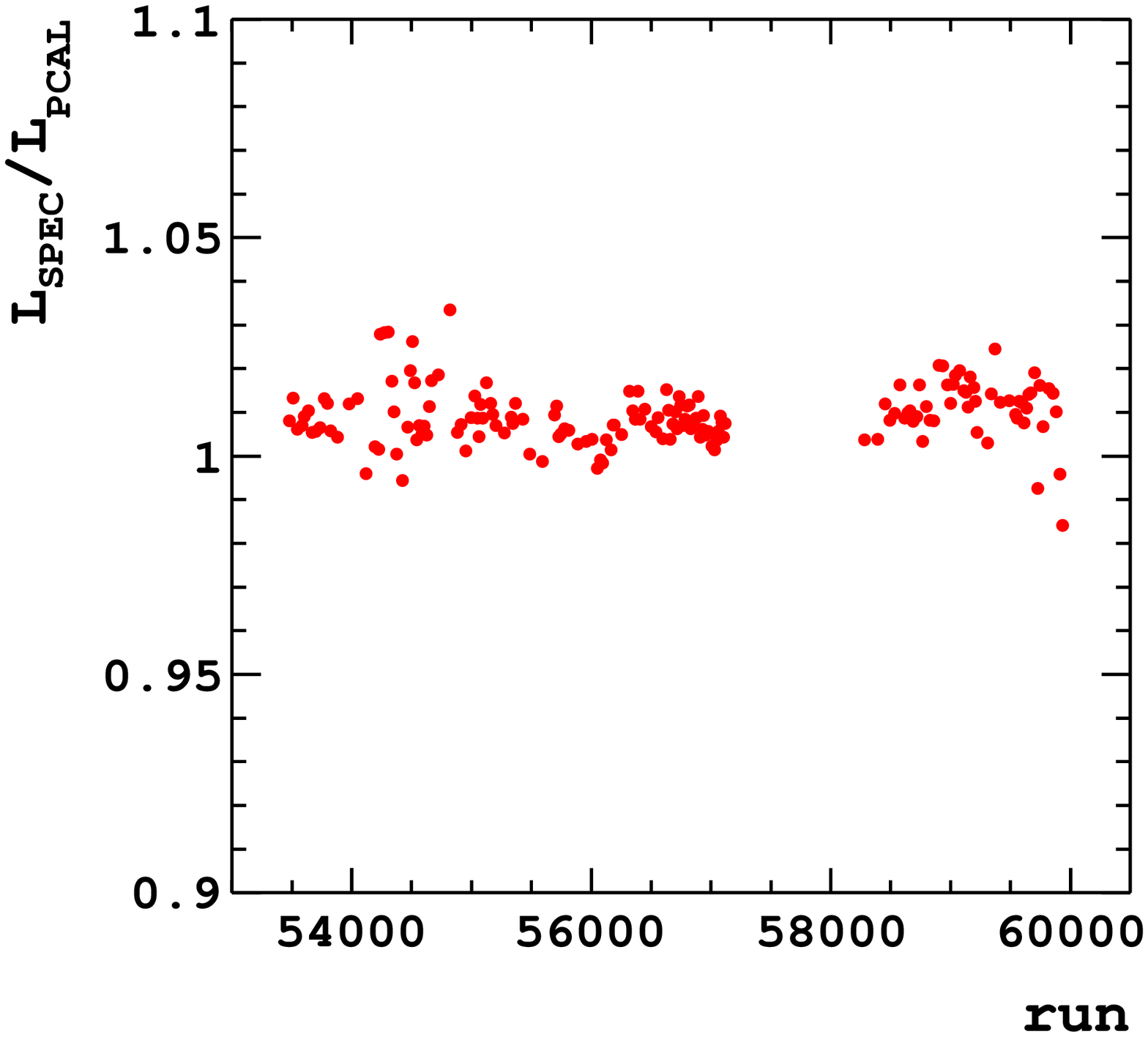}
\end{tabular}
\caption{ {Left:} The ratio of the luminosities measured by the
\spec{} and \pcal{} for each run, weighted with the luminosity 
measured using the \spec{}. Data from the 2005-2006
electron beam period 
are included.  {Right:} The ratio of the luminosities measured by the
\spec{} and \pcal{} plotted against the run number. Runs are grouped to 
include approximately 1\pbi per point.}
\label{fig-LLe-}
\end{figure}
\begin{figure}[!htb]
\begin{tabular}{cc}
\includegraphics[width=0.47\textwidth]{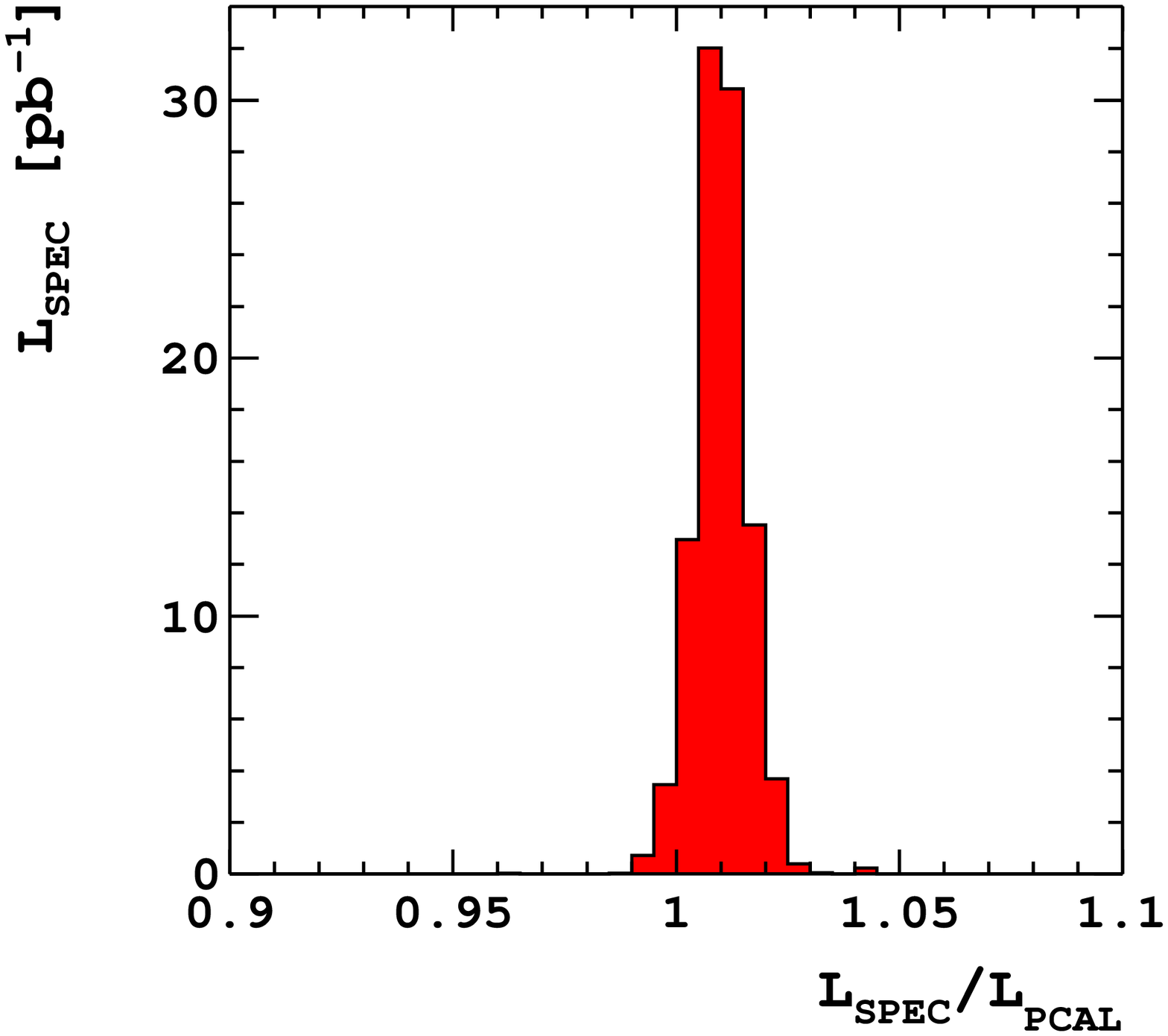}&
\includegraphics[width=0.47\textwidth]{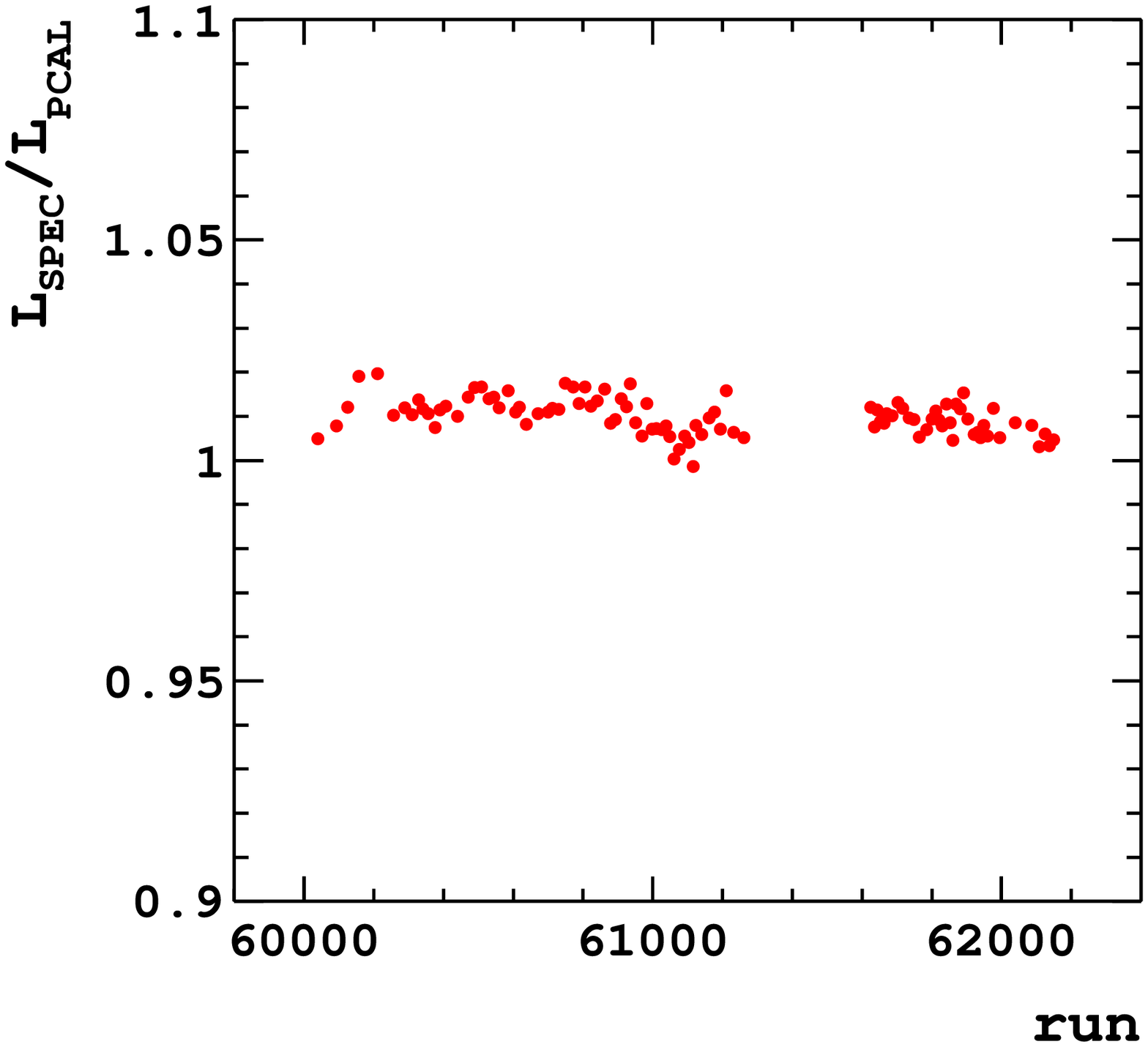}
\end{tabular}
\caption{ {Left:} The ratio of the luminosities measured by the
\spec{} and \pcal{} for each run, weighted with the luminosity 
measured using the \spec{}. Data from the 2006-2007
positron beam period 
are included.  {Right:} The ratio of the luminosities measured by the
\spec{} and \pcal{} plotted against the run number. Runs are grouped to 
include approximately 1\pbi per point.}
\label{fig-LLe+}
\end{figure}

During data taking the \spec~had a larger fraction of runs with malfunctions in the hardware.
However, the systematic uncertainties of the \spec~are smaller than the ones of the \pcal. 
Therefore, the luminosity
values of each run are taken from the \pcal~and 
multiplied with the ratio as shown in \Fig{s~\ref{fig-LLe-}}
and \ref{fig-LLe+} to obtain the luminosity values used in the physics analyses.
\section{Summary}

The luminosity in the ZEUS experiment was measured by counting photons from electron bremsstrahlung, $ep \to ep \gamma$.
The cross sections of this process is calculated in next-to-leading order in QED.  
Two independent devices, the \pcal~and the \spec, were used to measure the photon rate. The results of both 
agree within one per cent. 
The systematic uncertainty of the luminosity measurement is estimated to be 1.7\%.








\section*{Appendix}

To correct for the impact of pile-up the following 
model was used.
The mean number of electron pairs  
per bunch crossing, $\lambda$, with an energy larger than 8~\gev~reached 
up to 0.02 at the beginning of a fill. Neglecting the 
probability of more than 2 electron pairs per bunch crossing,
the fraction of colliding bunches with an 
event in the~\spec~passing all selection cuts, $r$, is
\begin{equation} \label{eq-r}
r = P_{0,1}A_{0,1} + P_{1,1}A_{1,1} + P_{0,2}A_{0,2},
\end{equation}
where 
$P_{0,1}$ and $A_{0,1}$ are the probability of the creation
and acceptance of an electron pair when no electron pair was in 
the previous bunch crossing and
$P_{1,1}$ and $A_{1,1}$ the same quantities when there was an
electron pair in the previous bunch crossing. 
$A_{1,1}$ accounts for losses due to 
previous bunch subtraction, which may lead to event 
rejection by the reduction of the deposited energy.
$P_{0,2}$ and $A_{0,2}$ are the probability and acceptance    
of events with two electron pairs 
which are counted as one event. 
$A_{0,2}$ accounts for fake coincidences made 
by $e^{+}$ and $e^{-}$ from different photons

The number of electron pairs per bunch crossing is described by a 
Poisson distribution with the mean $\lambda$. Evaluating 
probabilities $P_{m,n}$ to order $\lambda^{2}$ and solving 
\Eq{(\ref{eq-r})} for $\lambda(r)$ we have
\begin{equation} \label{eq-lambda}
\lambda = \frac{r}{A_{0,1}} + \left[ 1 - \frac{A_{0,2}}{2A_{0,1}} + f \left( 1 - \frac{A_{1,1}}{A_{0,1}}\right) \right] \left( \frac{r}{A_{0,1}} \right)^{2},
\end{equation}
where \emph{f} is the fraction of colliding bunches with 
a preceding colliding bunch. It accounts 
for the fact that the colliding bunches come in trains 
alternating with pilot and empty bunches.

The luminosity is
\begin{equation} \label{eq-ltrue}
L_{true} = N_{cb} \frac{\lambda(E_{\gamma} > E_{thres.})}{\sigma(E_{\gamma} > E_{thres.})},
\end{equation}
where $N_{cb}$ is the number of colliding bunches and $E_{thres.}$ is the energy threshold.

The measured luminosity is
\begin{equation} \label{eq-lmeas}
L_{measured} = N_{bc} \frac{r/A_{0,1}(E_{\gamma} > E_{thres.})}{\sigma(E_{\gamma} > E_{thres.})}.
\end{equation}

The relative luminosity shift is then
\begin{equation} \label{eq-dl}
\frac{L_{true}-L_{measured}}{L_{measured}} = \left[ 1 - \frac{A_{0,2}}{2A_{0,1}} + f \left( 1 - \frac{A_{1,1}}{A_{0,1}}\right) \right] \frac{r}{A_{0,1}}.
\end{equation}

The quantities $A_{m,n}$ are determined using Monte Carlo 
simulations of the photon beam with parameters obtained from data. 
Values of \emph{f} are calculated for the bunch sequence 
of each run.

\end{document}